# Cross-species analysis traces adaptation of Rubisco towards optimality in a low dimensional landscape

**Yonatan Savir[1], Elad Noor[2], Ron Milo[2] and Tsvi Tlusty[1*]**

*[1]Department of Physics of Complex Systems, [2]Department of Plant sciences, Weizmann Institute of Science, Rehovot 76100, Israel*

*[*]To whom correspondence should be addressed. E-mail: tsvi.tlusty@weizmann.ac.il*

Rubisco, probably the most abundant protein in the biosphere, performs an essential part in the process of carbon fixation through photosynthesis thus facilitating life on earth. Despite the significant effect that Rubisco has on the fitness of plants and other photosynthetic organisms, this enzyme is known to have a remarkably low catalytic rate and a tendency to confuse its substrate, carbon dioxide, with oxygen. This apparent inefficiency is puzzling and raises questions regarding the roles of evolution versus biochemical constraints in shaping Rubisco. Here we examine these questions by analyzing the measured kinetic parameters of Rubisco from various organisms in various environments. The analysis presented here suggests that the evolution of Rubisco is confined to an effectively one-dimensional landscape, which is manifested in simple power law correlations between its kinetic parameters. Within this one dimensional landscape, which may represent biochemical and structural constraints, Rubisco appears to be tuned to the intracellular environment in which it resides such that the net photosynthesis rate is nearly optimal. Our analysis indicates that the specificity of Rubisco is not the main determinant of its efficiency but rather the tradeoff between the carboxylation velocity and $CO_2$ affinity. As a result, the presence of oxygen has only moderate effect on the optimal performance of Rubisco, which is determined mostly by the local $CO_2$ concentration. Rubisco appears as an experimentally testable example for the evolution of proteins subject both to strong selection pressure and to biochemical constraints which strongly confine the evolutionary plasticity to a low dimensional landscape.

\body

## Introduction

Photosynthetic carbon assimilation enables the storage of energy in the global ecosystem and produces most of the global biomass. Rubisco (D-ribulose 1,5-bisphosphate carboxylase/oxygenase), probably the most abundant enzyme in nature (1), catalyzes the addition of $CO_2$ and $H_2O$ (2, 3) to 1,5-ribulose bisphosphate (RuBP) in the first major step of carbon fixation through photosynthesis. Rubisco is present in most autotrophic organisms from prokaryotes, such as photosynthetic anaerobic bacteria and cyanobacteria, to eukaryotes, such as algae and higher plants (4). The catalytic rate of Rubisco is remarkably slow. On top of that, Rubisco tends to catalyze the addition of $O_2$ instead of $CO_2$, leading to photorespiration which entails an extra energy investment and a reduction in the net photosynthetic rate (5). The seeming contradiction between the importance of Rubisco and its apparent inefficiency motivated an ongoing effort to improve Rubisco by genetic manipulation (6) and directed evolution (7, 8), with very limited success so far. One would desire to increase the specificity of Rubisco to $CO_2$ and its rate of carboxylation. However, this task proves difficult since the specificity and the carboxylation velocity appear to be negatively correlated (9). A possible biochemical mechanism for this tradeoff was recently proposed (10), following the hypothesis that Rubisco optimizes this tradeoff according to its environment. The present work quantitatively addresses the questions regarding the optimality and inefficiency of Rubisco, which we suggest result from an interplay between constraints and evolutionary forces. Our analysis delineates a low dimensional landscape shaped by underlying physico-chemical constraints in which Rubsico evolves. Comparison of cross species data implies that Rubisco is nearly optimal in this constrained landscape.

Carbon fixation by Rubisco is a multistage process (Fig. 1A) (11, 12). In the first stage, Rubisco binds RuBP and the formed complex undergoes enolization. This is followed by an irreversible $CO_2$ addition (carboxylation) which results in a six-carbon intermediate. Then, steps of hydration and cleavage yield two molecules of a three-carbon compound, 3-phosphoglycerate, which are later used to make larger carbohydrates. In the competing reaction of oxygenation, the Rubisco-RuBP complex irreversibly captures $O_2$ and through similar steps of hydration and cleavage yields only one 3-phosphoglycerate molecule and one molecule of 2-phosphoglycolate. In order to retrieve the carbons in 2-phosphoglycolate a complicated process of photorespiration takes place incurring a net loss of $CO_2$ (about one $CO_2$ molecule per two captured $O_2$ molecules (5)) and thus reduces the photosynthetic carboxylation rate.

The Rubisco-catalyzed carboxylation and oxygenation are known to exhibit *effective* Michaelis-Menten (MM) kinetics (Fig. 1B) (13). The MM parameters for carboxylation are the maximal carboxylation velocity $v_C$, which combines the steps of hydration and cleavage, and the MM constant for $CO_2$ addition, $K_C$, which represents the effective affinity of the carbon dioxide molecule to the enolized Rubisco-RuBP complex. The carboxylation rate per Rubisco molecule, $R_C$, when RuBP is in saturation, takes the familiar MM form, $R_C = v_C /(1 + K_C/[CO_2] + (K_C/K_O)([O_2]/[CO_2]))$. Addition of oxygen sequesters a fraction of the available Rubisco-RuBP complexes and is represented by the factor $[O_2]/K_O$ in the denominator, where $K_O$ is the effective MM constant for oxygen binding. A similar expression is derived for the rate of oxygenation per Rubisco molecule, $R_O$ (Fig. 1B). The specificity of Rubisco, $S$, is the ratio of the normalized carboxylation and oxygenation rates, $S = (R_C/R_O)\cdot([O_2]/[CO_2]) = (v_C/K_C)/(v_O/K_O) = k_{on,C}/k_{on,O}$, which depends solely on the ratio of the addition rates. The $CO_2$ and $O_2$ concentrations which affect the kinetic parameters are not determined solely by the ambient habitat conditions. Many species developed $CO_2$ concentrating mechanisms (CCM) that enable the accumulation of $CO_2$ at the carboxylation site (14-16).

It has been suggested that the oxygenation, which leads to the low specificity of Rubisco, is an inherent side effect of biochemical constraints on the reaction (2, 3, 17, 18). However, the specificity $S$ and other kinetic parameters do vary among species (9), which implies that selection pressure may play a role in shaping Rubisco in response to environmental changes. Correlations among the kinetic parameters from various organisms, in particular, the negative $S - v_C$ correlation (9, 19, 20) provide evidence for an interplay between constraints and selection and support an underlying structural mechanism (10). For example, Rubisco that has adapted to low $CO_2/O_2$ ratios tends to have high specificity but, on the other hand, low $v_C$.

## Results

### The kinetic parameters of Rubisco are confined to an effectively 1D landscape

In the following, we examine the interplay between the biochemical constraints and the evolutionary selection pressure to optimize Rubisco by analyzing the correlations among the measured *in vitro* kinetic parameters ($v_C$, $K_C$, $S$, $K_O$) from various organisms. Figure 2A presents the kinetic parameters of 28 Rubisco collected from 27 species (SI, Table S1), both eukaryotes and prokaryotes, which are divided into six groups, photosynthetic bacteria, cyanobacteria, green and non-green algae and $C_3/C_4$ higher plants. A few forms of Rubisco are known; form-I ($L_8S_8$) is composed of eight large and eight small subunits whereas form-II consists of only the large subunits (4). The Rubisco we had information for and analyzed here are all of the more abundant form-I except for the form-II Rubisco of *Rhodospirillum rubrum* and *Rhodopseudomonos sphaeroides*.

Each species in the data set corresponds to a point in a four-dimensional space whose coordinates are the four kinetic parameters: $v_C$, $K_C$, $S$ and $K_O$. When the data points are plotted in logarithmic scale (Fig. 2B) they appear to follow a straight line, which indicates that the data resides in an effectively one-dimensional space. To quantify this observation of reduced dimensionality, we performed principal component analysis (PCA) of the data (21). The PCA amounts to rotating the coordinate-system such that as much as possible of the variability in the data lies along one axis called the first principal component, which is actually the straight line that best represents the data in terms of least squares.

We analyze the data in terms of the four parameters which determine the rates of carboxylation and oxygenation, that is $v_C$, $K_C$, the ratio $K_C/K_O$ and $S$. Surprisingly, we find that the first principal component captures about 91% of the variability in the data (Methods) and therefore the data is indeed effectively one-dimensional. The data exhibit strong power-law correlations (Fig. 2) (two significant digits are shown),

$$K_C \propto v_C^{2.0 \pm 0.2}$$
$$\frac{K_C}{K_O} \propto v_C^{1.5 \pm 0.2} \qquad\qquad (1)$$
$$S \propto v_C^{-0.51 \pm 0.1}$$

.

There are two evident outliers, the only form-II Rubisco of *R. rubrum* and *R. sphaeroides,* which were therefore excluded from the fit. We also tested the possibility of separate power law correlations for prokaryotes and eukaryotes (SI, Fig. S1) However, the analysis shows that the form I prokaryotes follow almost the same trends as form I eukaryotes thus suggesting that the more relevant division is into form I and form II Rubisco which are subject to different constraints. The dependence of $K_C$, $S$, and $K_O/K_C$ solely on $v_C$ signifies the effective one-dimensionality of the data. The extracted power laws manifest inherent tradeoffs between the kinetic parameters of Rubisco. For example, the specificity $S$ decreases like the square root of the carboxylation velocity $v_C$.

## The carboxylation and oxygenation energetic tradeoffs

The effective MM kinetics can be represented in terms of a free energy profile (Fig. 3A) and the power law correlations between the kinetic parameters can be translated into energetic tradeoffs. The effective kinetics consists of two irreversible steps, effective gas binding (i.e., enolization and gas addition) and effective catalysis (i.e. hydrolysis and cleavage) (Fig. 1B), which correspond to two effective energy barriers. The values of the first energy barrier, $\Delta G_{1,C}$ for carboxylation and $\Delta G_{1,O}$ for oxygenation, are related to the effective gas binding rates, $k_{on,C} \sim \exp(-\Delta G_{1,C})$, $k_{on,O} \sim \exp(-\Delta G_{1,O})$. The second energy barriers, $\Delta G_{2,C}$ and $\Delta G_{2,O}$, are linked to the effective catalysis rates, $v_C \sim \exp(-\Delta G_{2,C})$, $v_O \sim \exp(-\Delta G_{2,O})$. The specificity $S$ is simply the exponent of the difference in the energy barrier for $CO_2$ and $O_2$ addition, $S \sim \exp(\Delta G_{1,O} - \Delta G_{1,C})$.

From the phenomenological power law relations (Eq. 1) we deduce the interplay between changes in the energy barriers across species. We find two basic tradeoffs: the first tradeoff follows from the power law $v_C/K_C = k_{on,C} \sim 1/v_C$ (Fig. 3B). This quantitatively manifests a previously suggested tradeoff between the addition of $CO_2$ to the enoled RuBP and the catalysis rate of this complex (10). This inverse relation can be expressed, by taking the logarithm, as the "conservation" of the sum of the carboxylation energy barriers, $\Delta G_{1,C} + \Delta G_{2,C} \approx$ const. (Fig. 3B). One may speculate that the origin of this tradeoff is the partition of a certain approximately constant deformation energy, which is required for the completion of the carboxylation process, into two sequential steps. A second tradeoff is between the $CO_2$ and $O_2$ addition rates, $k_{on,O} \sim k_{on,C}^{0.5}$, which indicates that a decrease in the $CO_2$ addition barrier is associated with a smaller decrease, by a factor of ½, in the $O_2$ addition barrier (Fig. 3C) such that $0.5 \cdot \Delta G_{1,C} - \Delta G_{2,O} \approx$ const. These two basic tradeoffs can be combined into the apparent tradeoff between specificity and carboxylation velocity, $S = k_{on,C} / k_{on,O} \sim v_C^{-0.5}$. An increase in the specificity is not due to a lower $O_2$ binding rate, $k_{on,O}$, which actually increases, but due to an even faster increase of the $CO_2$ binding rate, $k_{on,C}$. This resembles the recently suggested conformational proofreading mechanism (22) in which conformational changes – in the case of Rubisco, possibly linked to the closure of loop 6 (23) – simultaneously vary the rates of two competing reactions such that the overall specificity increases. No significant correlation is observed between the effective $O_2$ catalysis rate $v_O$ and the rest of the parameters (SI, Fig. S2). This hints that this reaction stage is only weakly coupled to the main tradeoffs and thus $v_O$ values can be selected for independently. Another possibility is that due to the relatively low $O_2$ concentrations, which are rarely above $K_O$ (Fig. 2A), the strongly selected quantity is $k_{on,O} = v_O/K_O$ and not $v_O$ itself.

### Adaptation of Rubisco in the 1D landscape is nearly optimal

A longstanding question is the optimality of Rubisco to its environment. The correlations between the kinetic parameters suggest the existence of underlying biochemical and structural limitations which constrain the evolution of Rubisco. The power law relations between the kinetic parameters of Rubisco depict an effectively one dimensional landscape which allows us to quantitatively examine the optimality of Rubisco under these "design constraints". As a measure of the fitness of Rubisco we use the net photosynthesis rate (NPR) per Rubisco molecule, $f$, which is the difference between the fixed $CO_2$ and the $CO_2$ lost due to oxygenation (24). For each fixed mole of $O_2$ molecule about $t \approx$ ½ moles of $CO_2$ are lost and the NPR is therefore $f = R_C - t \cdot R_O$. The NPR depends on the four parameters, $v_C$, $K_C$, $S$, and $K_C/K_O$. However, these parameters are not independent and thus the NPR is a 1D landscape determined by one kinetic parameter, for example $v_C$,

$$f = \frac{v_C - 3 \cdot 10^{-3} \cdot ([O_2]/[CO_2]) \cdot v_C^{3/2}}{1 + 1.3 \cdot v_C^2 / [CO_2] + 5 \cdot 10^{-3} \cdot ([O_2]/[CO_2]) \cdot v_C^{3/2}}, \tag{2}$$

where the concentrations are given in μM and $v_C$ in units of 1/sec. The NPR in a given intracellular environment exhibits a clear maximum as a function of $v_C$. For example, Rubisco of $C_3$ plants, which lack a CCM, operate at $[CO_2]$ of around 7-8 μM, while Rubsico of $C_4$ plants, which have a CCM, experience $CO_2$ concentrations which are at least ten times larger (25). Comparison of the optimal $v_C$ to the measured ones (Fig. 4A) indicates that the Rubisco from $C_4$ plants is nearly optimal at $[CO_2] = 80$ μM whereas $C_3$ plants are too slow for this environment.

In the absence of oxygen ($[O_2] = 0$) the NPR is simply the carboxylation rate, $f = R_C = v_C/(1 + K_C/[CO_2]) = v_C/(1 + 1.3 \cdot v_C^2/[CO_2])$. The evident optimum is the direct outcome of the tradeoff between the $CO_2$ affinity and the carboxylation velocity, $K_C \sim v_C{}^2$. At the limit of low $v_C$ , $K_C$ is also low ($K_C$ << [$CO_2$]) and the enzyme is in saturation. Thus, $f$ increases linearly with $v_C$ , $f \sim v_C$. At the other extreme of high $v_C$ and $K_C$ ($K_C$ >> [$CO_2$]) the NPR decreases due to the fact that the affinity increases faster than the velocity and thus, $f \sim 1/v_C$. We find simple expressions for the optimal values of the kinetic parameters that bring $f$ to its optimum $f^*$ in an anaerobic environment

$$v_C^* \cong 0.86 \cdot [\text{CO}_2]^{1/2}$$
$$K_C^* \cong [\text{CO}_2]$$
$$S^* \cong 164 \cdot [\text{CO}_2]^{-1/4}$$

(3)

The resulting optimal NPR is about half of the maximal carboxylation velocity, $f^* \sim v_C^*/2 \sim 0.45 \cdot [\text{CO}_2]^{1/2}$ s$^{-1}$. The coefficient of the specificity power law of about 200 indicates a difference between the gas addition barriers of about 3-5 $k_B T$ which coincides with previous estimations (23).

In aerobic environments, the presence of oxygen reduces the net photosynthesis in two ways (Fig. 4A, 4B), sequestering a fraction of the available Rubisco, akin to competitive inhibition, and the loss of $CO_2$ due to $O_2$ fixation, which reduces $f$ by a factor of $t \cdot R_O$. Because of these two effects, the presence of oxygen shifts the optimal carboxylation velocity $v_C^*$ to lower values thus improving the specificity and reducing oxygen addition. For example, for [$CO_2$] = 80 μM, oxygen levels of 260 μM lead to reduction of around 17% in NPR and 7% in $v_C^*$ due to Rubisco sequestering whereas another 3% reduction in NPR and 1% in $v_C^*$ is due to $O_2$ fixation (Fig. 4A). It is evident that the optimal value of the carboxylation velocity $v_C^*$ is dictated by the concentration of $CO_2$ and both effects of oxygen are only smaller corrections. Approximate expressions that include the effect of oxygen on the optimal kinetic parameters are

$$v_C^* = 0.86 \cdot [\text{CO}_2]^{1/2} \left(1 - \delta_O \right)$$
$$K_C^* = [\text{CO}_2] \left(1 - 2 \cdot \delta_O \right)$$
$$S_C^* = 164 \cdot [\text{CO}_2]^{-1/4} \left(1 + 0.5 \cdot \delta_O \right).$$

(4)

The small parameter $\delta_O = 10^{-3} \cdot [\text{O}_2][\text{CO}_2]^{-1/4}$ accounts for the effect of oxygen, which for all relevant conditions ranges between 0-15%. In other words, the specificity of Rubisco is not the main determinant of its efficiency but rather the $v_C$-$K_C$ tradeoff between the carboxylation velocity and affinity is the dominant effect. The resulting optimal NPR is reduced by the presence of oxygen as, $f^* = 0.45 \cdot [\text{CO}_2]^{1/2} (1 - 2 \cdot \delta_O)$ and the reduction ranges between 0-30%.

To further examine the optimality of Rubisco, we have to consider the intracellular environments of various organisms which are characterized by the concentrations of carbon dioxide and oxygen at the carboxylation site.

Many species have the capacity to increase the local concentration of $CO_2$ above the passive concentration by carbon dioxide concentration mechanisms (CCM). Therefore, besides the environment of $CO_2 = 80\mu M$ discussed above, which corresponds to medium range CCM (some $C_4$ plants, some algae, anaerobic bacteria) (*15, 16, 25-28*), we also plot the net photosynthesis rate for two other typical carbon dioxide concentrations (Fig. 4B): $[CO_2] = 10$ $\mu M$, which corresponds to the groups of no CCM ($C_3$ plants and some algae) (14-16, 28), and $[CO_2] = 250\mu M$, which corresponds to strong CCM (cyanobacteria) (15, 16, 28, 29). As there is only limited knowledge of the accurate values of $[CO_2]$ and $[O_2]$ for each species, the values we use are merely gross estimates which represent coarse classification into three typical environments.

For all classes of species, we find that the observed carboxylation velocity and $CO_2$ effective binding affinity are close to the optimal values, $v_C^*$ and $K_C^*$. This is also demonstrated in figure 4C which shows the optimal environment, $[CO_2]^*$, that corresponds to a measured carboxylation velocity. We find that the cyanobacteria is optimal in $CO_2$-rich environment, $[CO_2]^* \approx 240\mu M$, whereas $C_4$ plants, algae and photosynthetic bacteria are optimal in intermediate $CO_2$ levels, $[CO_2]^* = \approx 30\text{-}80\mu M$. Finally, $C_3$ plants and non-green algae which are suspected to lack CCM are optimal at $CO_2$ levels of 5-15 $\mu M$.

As a measure for the performance of Rubisco it is instructive to look at a landscape of the normalized NPR (Figure 5A). A value of unity means that the Rubisco performs at the maximal possible NPR for the given environment. For example, the cyanobacteria in a strong CCM environment, $[CO_2] = 250$ $\mu M$, has normalized NPR of almost 100%. However, if one had taken this Rubisco and put it in an environment typical of $C_3$ plants, then it would have achieved only 32% of the maximal possible NPR in this environment. In accord, photosynthesis was impaired when Rubisco of *R. rubrum* replaced the native Rubisco of cyanobacteria or tobacco (30, 31).

## Discussion

Several scenarios could lead to the observed effectively one-dimensional landscape (Fig. 5B): The constraints may be strict and all feasible kinetic parameters are therefore close to the line. In another scenario, the constraints are only upper limits on the kinetic parameters and it is selection that pushes Rubisco to this limit. In both cases, the resulting observed landscape is the 1D line, but the two scenarios differ in the accessible phenotypes. In the first, mutations cannot yield phenotypes far from the line, whereas in the latter, phenotypes far from the line are feasible but are expected to vanish rapidly by the strong selective forces. A hint that supports the latter scenario comes from the observed fluctuations of the kinetic parameters around the line. Combinations of parameters that strongly affect the NPR tend to exhibit much smaller variability. For example, the NPR does not depend directly on $K_O$ but rather on the on the affinity ratio $K_C/K_O$ (Fig. 1). Indeed, the correlation coefficient between $K_O$ and $v_C$ is about 0.5, indicating large variability, whereas the correlation coefficient between $K_C$ and $v_C$ is much larger, 0.95. A possible experimental test that could map the accessible phenotypes is a statistical survey of Rubisco phenotypes (i.e., their kinetic parameters) resulting from point mutations.

The fact that the only outliers are the form-II Rubisco of *R. rubrum* and *R. sphaeroides*, which lack the small subunits of form-I, may indicate that the two forms of Rubisco may be subject to different constraints. Measurements of the activity of isolated large subunits, especially from the Rubisco of *Synechococcus* (32-35), indicate that the $v_C$ is drastically reduced whereas the specificity is relatively unchanged. However, the main deviation of the form-II Rubisco from the correlations is in the specificity *S* (Fig 2B,C) whereas its $K_C$ and $v_C$, which are the main determinants of the NPR, obey the same power law correlations of the form-I Rubisco. This may hint that this tradeoff is linked to the large subunit, whereas the tradeoff between $k_{on,C}$ and $k_{on,O}$ is related to the small subunit. Measurements of other form-II Rubisco may further clarify the origin of this deviation.

Our analysis yields simple quantitative predictions for the response of Rubisco to changing environments (Eqs. (4)), which can be experimentally tested. For example, one may vary the ambient $CO_2$ levels, or alternatively manipulate the CCM in order to affect the concentration at the carboxylation site. We expect that the kinetic parameters of Rubisco will adapt to the change in order to optimize its performance. Another possibility is to replace the Rubisco of one species by a heterologous Rubisco from a species that lives in a different environment and trace its adaptation to the environment of the host organism. We note that the NPR, *f*, measures the effect of Rubisco on fitness and that the overall fitness of a species should take into account resources (light, water etc.). However, our conclusion regarding the one-dimensional landscape of Rubisco does not depend on the measure of optimality.

Our results indicate that Rubisco is close to optimality in the NPR and therefore cannot be significantly improved by point mutations. To improve the performance of Rubisco one may perhaps focus on improving the CCM rather than mutating the Rubisco itself. Nevertheless, the results do not preclude the possibility that a drastic change, such as the change between forms I and II, may result in Rubisco that is subject to different constraints, which may perhaps allow better performance.

Here we show that Rubisco sets an example for a protein whose plasticity is confined to a low dimensional landscape by underlying constraints. In this confined landscape, selection forces drive it towards optimality. It demonstrates how the interplay between selection and constraints limits the plasticity of proteins and their ability to explore the phenotype space in order to improve their fitness. Similar tradeoffs may shape the evolution of other multi-stage enzymes.

## Methods

***The effective Michaelis-Menten parameters.*** The $CO_2/O_2$ addition to the enolized Rubisco-RuBP complex is practically irreversible, $k'_{a,C}$, $k'_{a,O}$ $\approx 0$ (12, 36) and thus the apparent Michaelis-Menten constants for $CO_2$ addition ($K_C$) and $O_2$ addition ($K_O$) are (Fig. 1B): $K_C = (v_C/k_{a,C})\cdot(k_e + k'_e)/k_e$ and $K_O = (v_O/k_{a,O})\cdot(k_e + k'_e)/k_e$. There is as an uncertainty as to whether the addition of gas and water are sequential or concerted (12). In the sequential case, $CO_2$ addition results in a six-carbon carboxyketone intermediate which is followed by a gem-diol hydrate intermediate which, in turn, undergoes cleavage. In the concerted case, there is only a hydrated intermediate, which undergoes cleavage. Experimental data suggest that the six-carbon intermediate is stabilized on enzyme mostly in its hydrate form and imply that hydration may be irreversible (37). At physiological pH values, enolization should be faster than the maximal catalytic rate, $v_C$, $v_O << k_e$ (38, 39). Thus, the maximal rate of carboxylation ($v_C$) and oxygenation ($v_O$) take the form, $v_C = k_{cle,C}\cdot k_{hC}/(k_{cle,C} + k_{hC})$ and $v_O = k_{cle,O}\cdot k_{hO}/(k_{cle,O} + k_{hO})$. If the gaseous addition and hydration are concerted, then $v_C = k_{cle,C}$ and $v_O = k_{cle,O}$.

***PCA analysis and Total Least Squares analysis.*** The data set contains 4 variables which determine the NPR ($K_C$, $v_C$, $S$, $K_C/K_O$), and represent points in a 4D space. For 28 Rubisco $S$ and $K_C$ are known. From these 28, for 25 Rubisco $K_O$ is known and from these 25, for 16 Rubisco $v_C$ is known. We performed PCA analysis on the data from the 16 Rubisco for which all four kinetic parameters are available excluding the form II *R.Rubrum* outlier. The eigenvalues of the covariance matrix (the latent vector) are [3.6, 0.25, 0.09, 0.02] and their proportions are [91%, 6%, 2%, 1%]. To account for the rest of the Rubiscos, we performed a total least squares fit on the entire set. We find the parameters which minimize the distance between the data points in logarithmic scale and the 1D line (which represents our model) ($t$, $\alpha t+\beta$, $\alpha_1 t+\beta_1$, $\alpha_2 t+\beta_2$). The results yield the following power laws:

$$
\begin{aligned}
K_C &= 1.32 \pm 0.5 \cdot v_C^{2.03\pm0.2} \quad [0.49 \ \ 2.54]; [1.68 \ \ 2.56] \\
S &= 152 \pm 28 \cdot v_C^{-0.51\pm0.1} \quad [111 \ \ 245]; [-0.76 \ \ -0.34] \\
K_O &= 239 \pm 50 \cdot v_C^{0.57\pm0.1} \quad [136 \ \ 347]; [0.37 \ \ 0.87]
\end{aligned}
\tag{5}
$$

The confidence bounds (95%) for the prefactor and the exponent are in square brackets. In the total least square process, we excluded two evident outliers, the form-II of *R. rubrum* and *R.sphaeroides*. (*R.sphaeroides* was not part of the PCA data).

## Figure legends

**Figure 1. Carboxylation/Oxygenation by Rubisco.** (**A**) Rubisco catalyzes the addition of $CO_2$ or $O_2$ to D-ribulose 1,5 bisphosphate (RuBP). Rubisco binds RuBP to form a complex that undergoes enolization. In the case of carboxylation (upper pathway), this is followed by practically irreversible $CO_2$ addition which results in a six-carbon intermediate. Through steps of hydration and cleavage, the reaction produces two molecules of 3-phosphoglycerate (3-PGA). Oxygenation follows similar steps (lower pathway), thorough five-carbon intermediate and results in one 3-PGA and one 2-phosphoglycolate (2-PGY) (12). (**B**) The Rubisco-catalyzed carboxylation and oxygenation exhibit effective Michaelis-Menten (MM) kinetics. The carboxylation rate per Rubisco molecule, $R_C$, and oxygenation rate per Rubisco molecule, $R_O$, when RuBP is in saturation, take the familiar MM form (see equations in the figure). The effective kinetics consists of two irreversible steps. The effective gas binding ($k_{on,C}$, $k_{on,O}$) which consists of the enolization of the Rubisco-RuBP complex, whose fraction is $r = k_e/(k'_e + k_e)$, and gas addition, determined by the rates $k_{a,C}$ and $k_{a,O}$. This is followed by effective catalysis ($v_C$, $v_O$), which includes the steps of hydration and cleavage. The MM constants for gas addition, $K_C$ and $K_O$, are the effective affinities of the $CO_2$ and $O_2$ molecules to the Rubisco-RuBP complex (see Methods).

**Figure 2. Power law correlations among kinetic parameters of Rubisco define 1D landscape.** (**A**) The four kinetic parameters of 28 Rubisco from 27 species (SI, Table S1); the effective Michaelis-Menten (MM) constant for $CO_2$ binding, $K_C$ [μM], the maximal carboxylation rate, $v_C$ [1/sec], the specificity, $S$, and the effective MM constant for $O_2$ binding, $K_O$ [μM]. The parameters $K_C$ and $S$ are known for all 28 Rubisco. For 25 Rubisco $K_O$ is known and for 16 Rubisco all the kinetic parameters are available. All Rubisco in the data set are of the more abundant form-I besides the form-II Rubisco from *R. rubrum* and *R. sphaeroides*. (**B**) **Top Left:** A 3D projection of the 4D kinetic parameter data. Data from the 16 species for which all 4 kinetic parameters are available are graphed in logarithmic scale. Each of the 16 species is represented by a point whose coordinates are its $v_C$, $K_C$ and $S$ (blue spheres). The plot depicts the PCA result that the data is confined to an effectively 1D space and follow, in logarithmic scale, a straight line (blue cylinder). The cylinder axis is the first principal axis and its radius is the standard deviation from this axis. There is one evident outlier (blue triangle), the Rubisco from *R. rubrum*, the only form-II Rubisco in this set. **Top Right and Bottom:** Total least squares fit of the data from all 28 Rubisco (Methods) to a 1D power law model (black lines). The fits are plotted in logarithmic scale and exhibit clear power law correlations between the kinetic parameters. The correlation coefficient, $\rho$, and the p-value, $p$, are shown. As before, there are clear outliers (black triangles) which correspond to the form-II species, *R. rubrum* and *R. sphaeroides*. Note that deviations of form-II species from the model are significant mostly in its projections onto the specificity parameter $S$.

**Figure 3. The free energy tradeoffs.** **(A)** Representation of the effective Michaelis-Menten (MM) kinetics in terms of free energy profiles. The effective kinetics consists of two irreversible steps, effective gas binding (i.e., enolization and gas addition) and effective catalysis (i.e., hydrolysis and cleavage) (Fig. 1B), which correspond to two effective energy barriers. The first energy barriers are related to the rates of effective gaseous addition, $k_{on,C} \sim \exp(-\Delta G_{1,C})$, $k_{on,O} \sim \exp(-\Delta G_{1,O})$. The second barriers are linked to the effective catalysis rates, $v_C \sim \exp(-\Delta G_{2,C})$, $v_O \sim \exp(-\Delta G_{2,O})$. The specificity depends on the difference between the first energy barriers, $S = \exp(\Delta G_{1,O} - \Delta G_{1,C})$. **(B)** The power law correlation $k_{on,C} = v_C/K_C \sim 1/v_C$ (solid line, black triangle is the *R.Rubrum* outlier), indicates that a shift in carboxylation barrier, $\Delta G_{2,C}$, is accompanied by a shift by the same magnitude in $CO_2$ addition barrier, such that their sum is conserved, $\Delta G_{1,C} + \Delta G_{2,C} = $ const. The change in $\Delta G_{2,C}$ in the figure involves only a shift in the intermediate energy level but may also involve a shift in the catalysis transition-state level. **(C)** Another tradeoff arises from the correlation between the effective addition rates of $CO_2$ and $O_2$, $k_{on,O} = v_O/K_O \sim (v_C/K_C)^{0.5} = k_{on,C}^{0.5}$ (solid line, black triangle is the *R.Rubrum* outlier). This indicates that a decrease in the $CO_2$ addition barrier, $\Delta G_{1,C}$, is associated with a smaller (by a factor of ½) decrease in the $O_2$ addition barrier, $\Delta G_{1,O}$, such that $0.5 \cdot \Delta G_{1,C} - \Delta G_{2,O} = $ const. The negative correlation between $v_C$ and $S$ is the outcome of the two tradeoffs: As $v_C$ decreases, $k_{on,C}$ increases and ,as a result, $k_{on,O}$ also increases, but by a smaller factor, and the difference between the gas addition barriers increases resulting in a decrease of the specificity, $S = \exp(\Delta G_{1,O} - \Delta G_{1,C})$.

**Figure 4. The optimality of Rubisco.** **(A)** Carboxylation rate, $R_C$, oxygenation rate, $R_O$, and net photosynthesis rate (NPR), $f = R_C - 0.5 \cdot R_O$, as a function of carboxylation velocity, $v_C$, for $[CO_2] = 80 \mu M$. In an anaerobic environment the NPR equals the carboxylation rate, $f = R_C$ ($[O_2] = 0$) (blue dashed line) and exhibits a clear optimum. The presence of oxygen, $[O_2] = 260 \mu M$, reduces the NPR (green line) in two ways: $R_C$ (blue line) is reduced since oxygenation sequesters a fraction of the available Rubisco, an effect which is responsible for most of the NPR decrease. A smaller reduction of the NPR is due to the photorespiration factor, $0.5 \cdot R_O$ (red line). The presence of oxygen shifts the optimal carboxylation velocity $v_C^*$ towards lower values (4). Most of the shift in $v_C^*$ is due to the decrease in $R_C$ rather than the increase in $R_O$. For example, $C_3$ plants which operate at $[CO_2] \approx$ 7-8$\mu M$, are far from the optimal values for $[CO_2] = 80 \mu M$. However, $C_4$ plants possess CCM and operate at $CO_2$ concentrations that are at least 10 times larger than those of $C_3$ plants and their carboxylation rates are nearly optimal at $[CO_2] = 80 \mu M$. **(B)** The NPR as a function of $v_C$ for three environments, $[CO_2] = 10$, 80 and 250 $\mu M$, which correspond to groups with no CCM ($C_3$ plants and some algae), medium range CCM ($C_4$ plants, some algae, photosynthetic bacteria) and strong CCM (cyanobacteria). The average $v_C$ of each class is plotted and appears to be close to the values which yield maximal NPR. The solid and dashed curves correspond to aerobic and anaerobic conditions, $[O_2] = 260$ and 0 $\mu M$, respectively. The optimal parameters of Rubisco are determined mostly by the $CO_2$ concentration. **(C)** The $CO_2$ environments that are predicted to be optimal to the observed $v_C$

from Eq. (4) (dashed line – anaerobic, solid line – $[O_2]$ = 260 μM). Cyanobacteria are optimal in $CO_2$-rich environment, $[CO_2]^* \approx 240$μM. $C_4$ plants, algae and photosynthetic bacteria are optimal at intermediate $CO_2$ levels, $[CO_2]^* \approx 30$-80μM. $C_3$ plants and non-green algae, which are suspected to lack CCM, are optimal at low $CO_2$ levels, $[CO_2]^* \approx 5$-15μM.

**Figure 5. (A)** Normalized NPR as a function of $v_C$ (fully optimal Rubisco has a normalized NPR of 100%). For example, the cyanobacteria in strong CCM environment (red line), $[CO_2]$ = 250 μM, has normalized NPR of about 99%, whereas it would have achieved only 34% of the maximal possible NPR in an environment experienced by $C_3$ plants, $[CO_2]$ = 10 μM. However, in their native environment, $C_3$ plants have normalized NPR of about 95%. **(B)** The axes represent any two kinetic parameters with negative correlation (dashed line), such as S and $v_C$. Rubisco (blue dots) reside in an effectively one-dimensional fitness landscape (the region near the correlation line), which may be the outcome of two possible scenarios: In the first, Rubisco is confined to a limited region of phenotypes (left). In the second, the observed relation may be only an upper limit (right) and it is selection that pushes Rubisco to the edge.

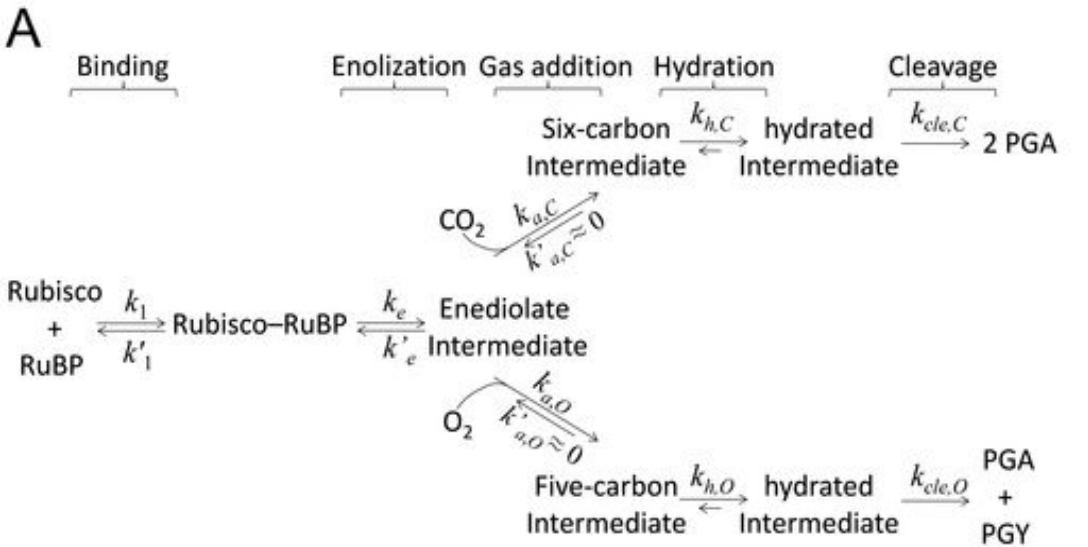

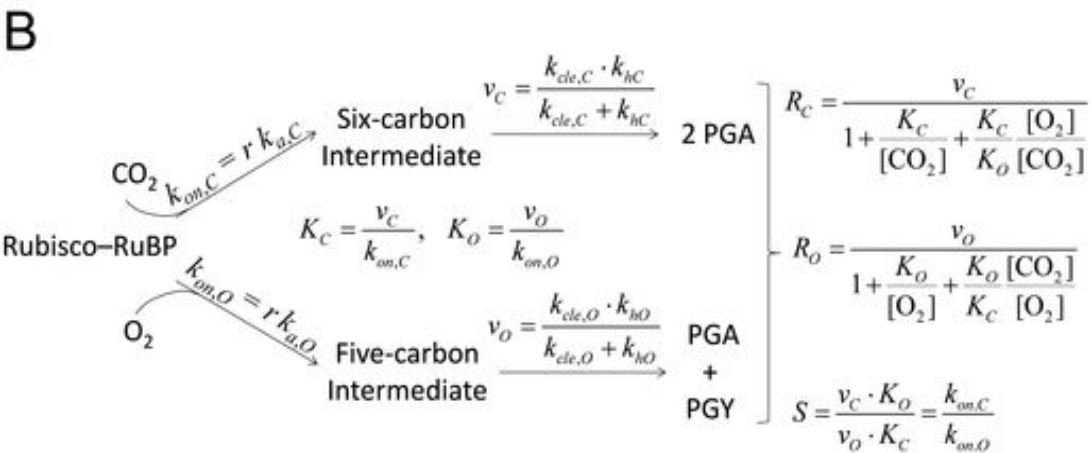

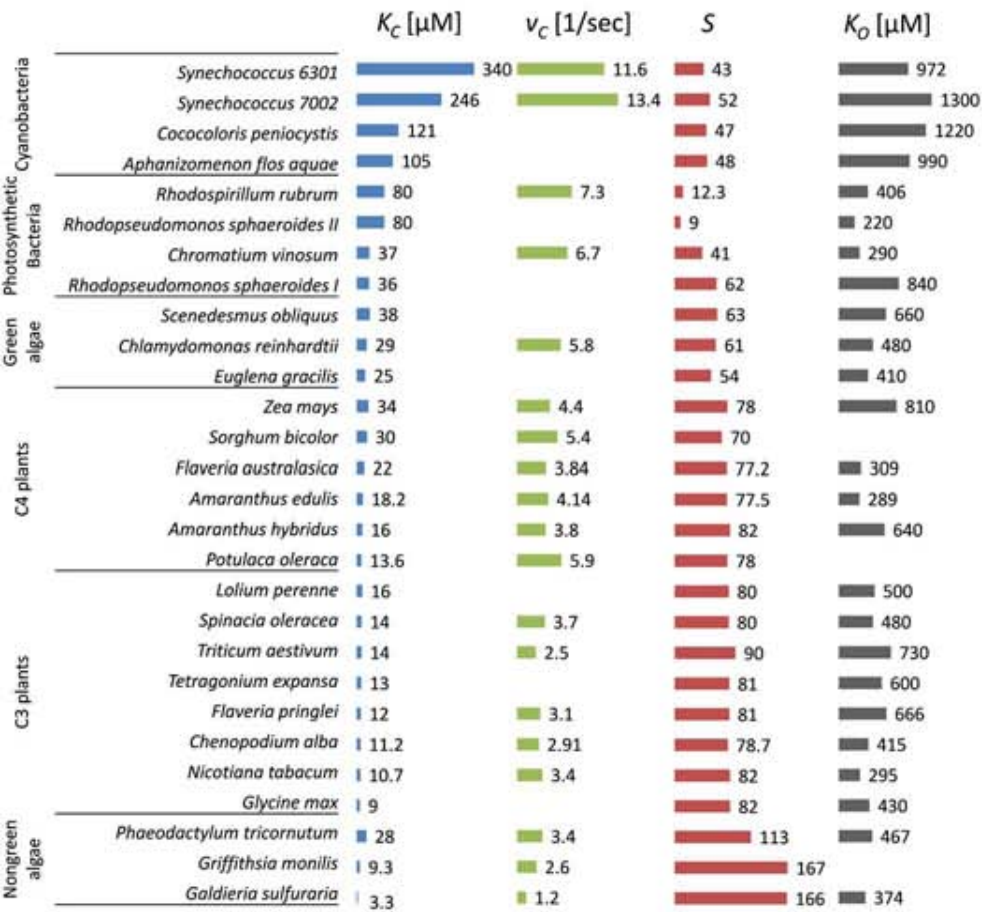
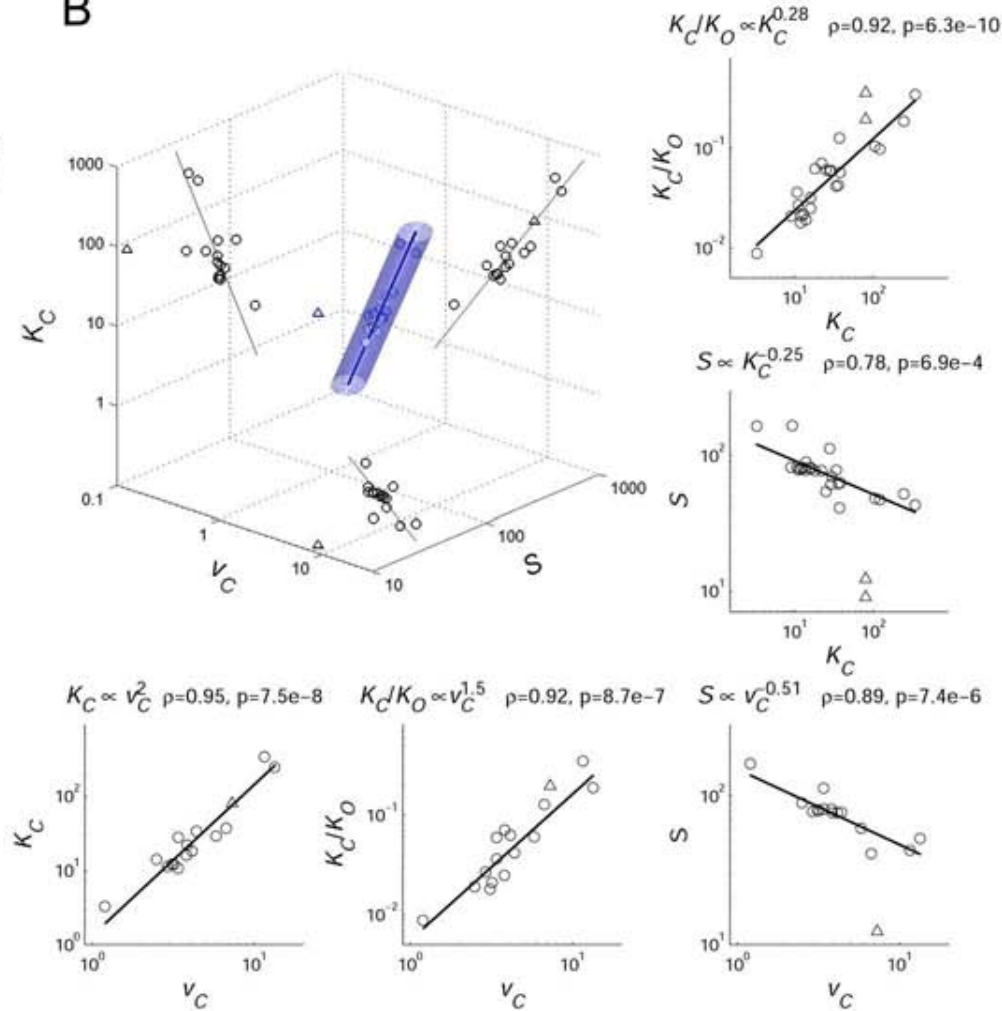

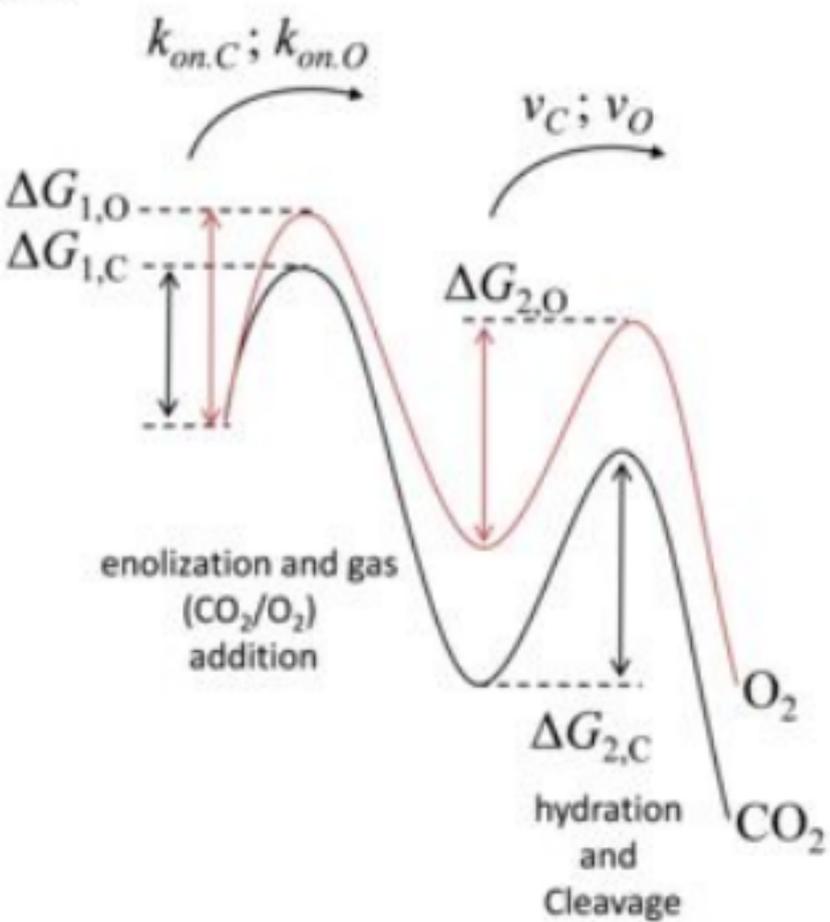

A

$k_{on,C}; k_{on,O}$

$v_C; v_O$

$\Delta G_{1,O}$
$\Delta G_{1,C}$

$\Delta G_{2,O}$

enolization and gas
($CO_2/O_2$)
addition

$\Delta G_{2,C}$

hydration and
Cleavage

$O_2$

$CO_2$

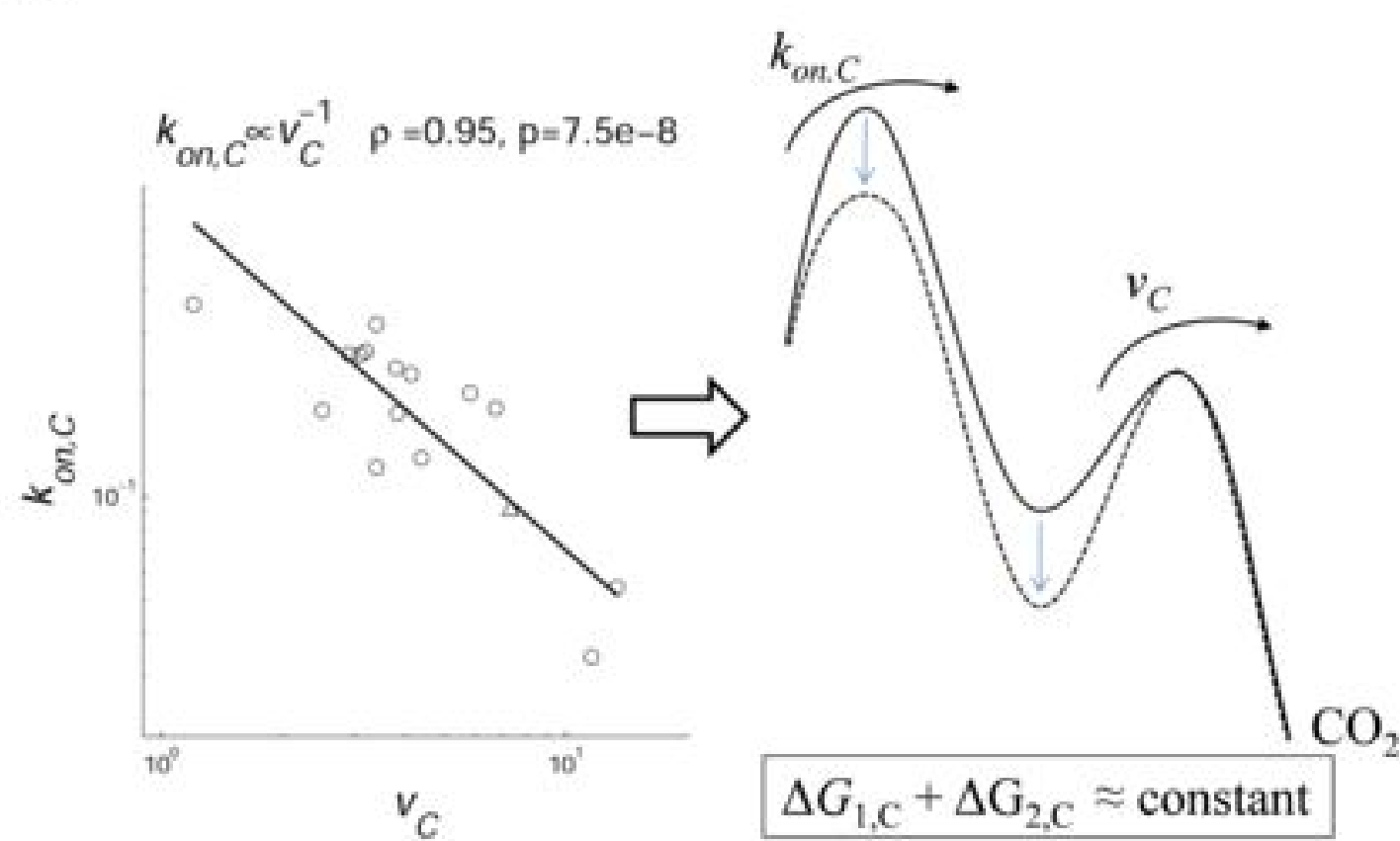

B

$k_{on,C} \propto v_C^{-1}$   $\rho = 0.95$, $p = 7.5\mathrm{e}{-8}$

$k_{on,C}$

$k_{on,C}$

$v_C$

$k_{on,C}$

$v_C$

$CO_2$

$\Delta G_{1,C} + \Delta G_{2,C} \approx \text{constant}$

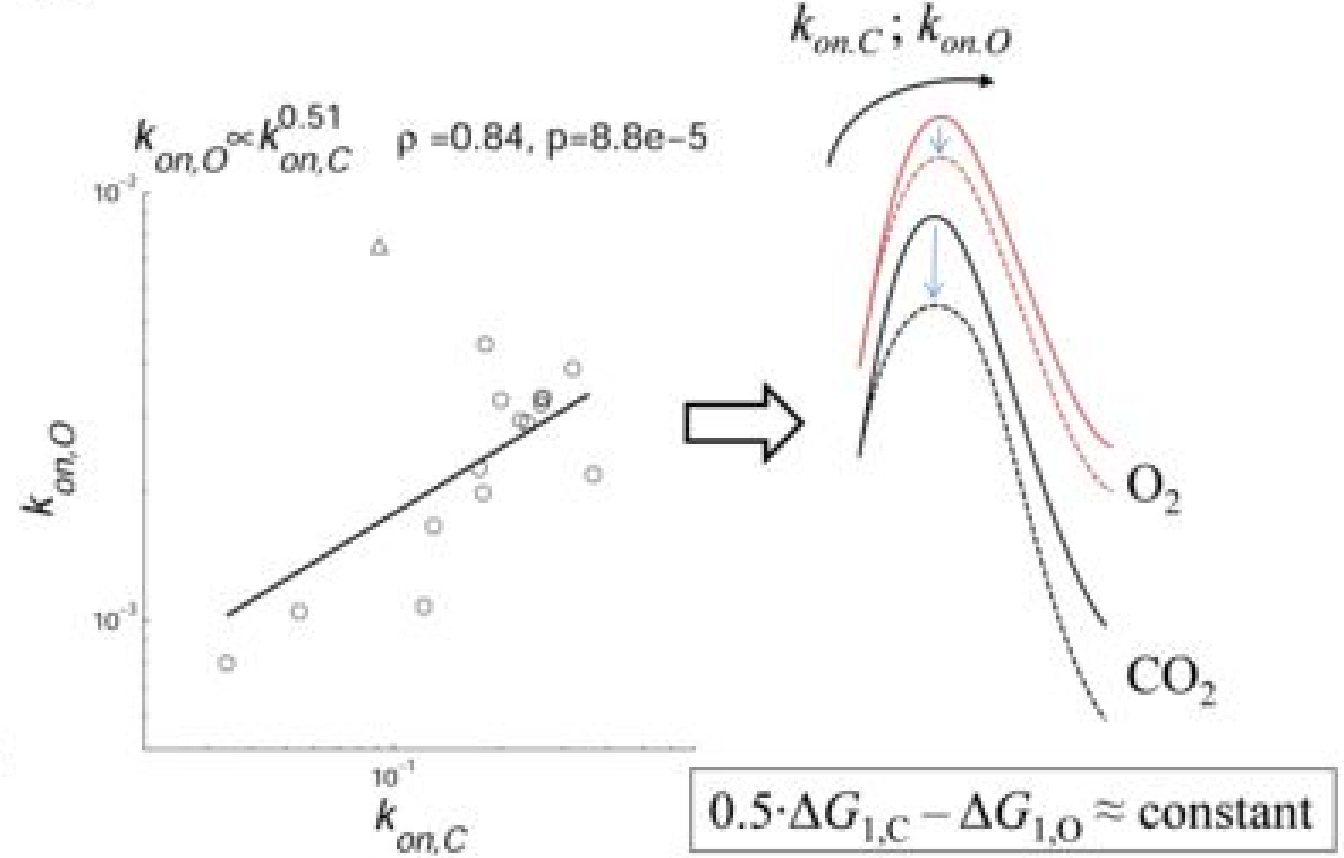

C

$k_{on,O} \propto k_{on,C}^{0.51}$   $\rho = 0.84$, $p = 8.8\mathrm{e}{-5}$

$k_{on,O}$

$k_{on,C}$

$k_{on,C}; k_{on,O}$

$O_2$

$CO_2$

$0.5 \cdot \Delta G_{1,C} - \Delta G_{1,O} \approx \text{constant}$

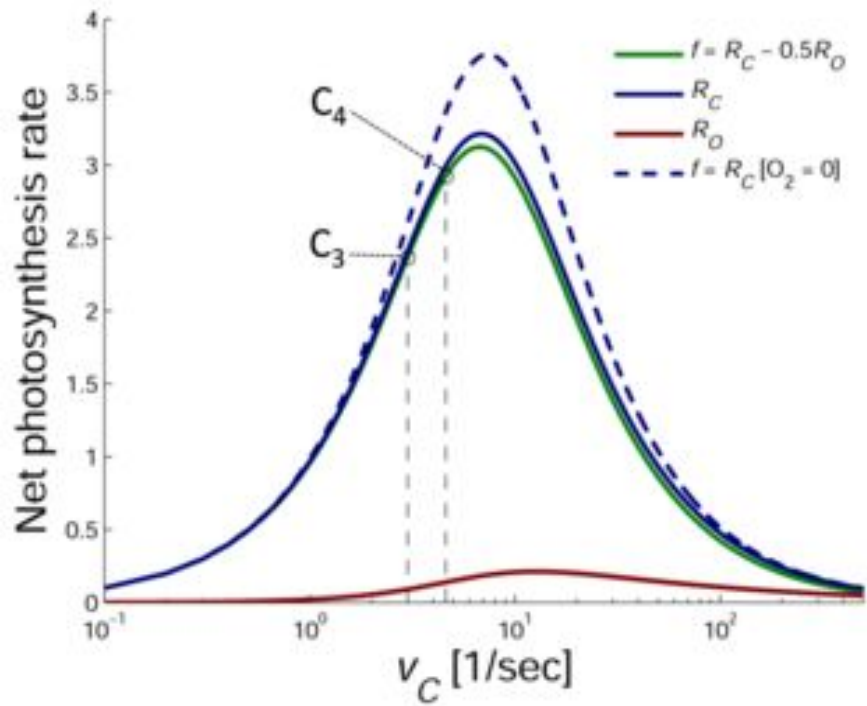

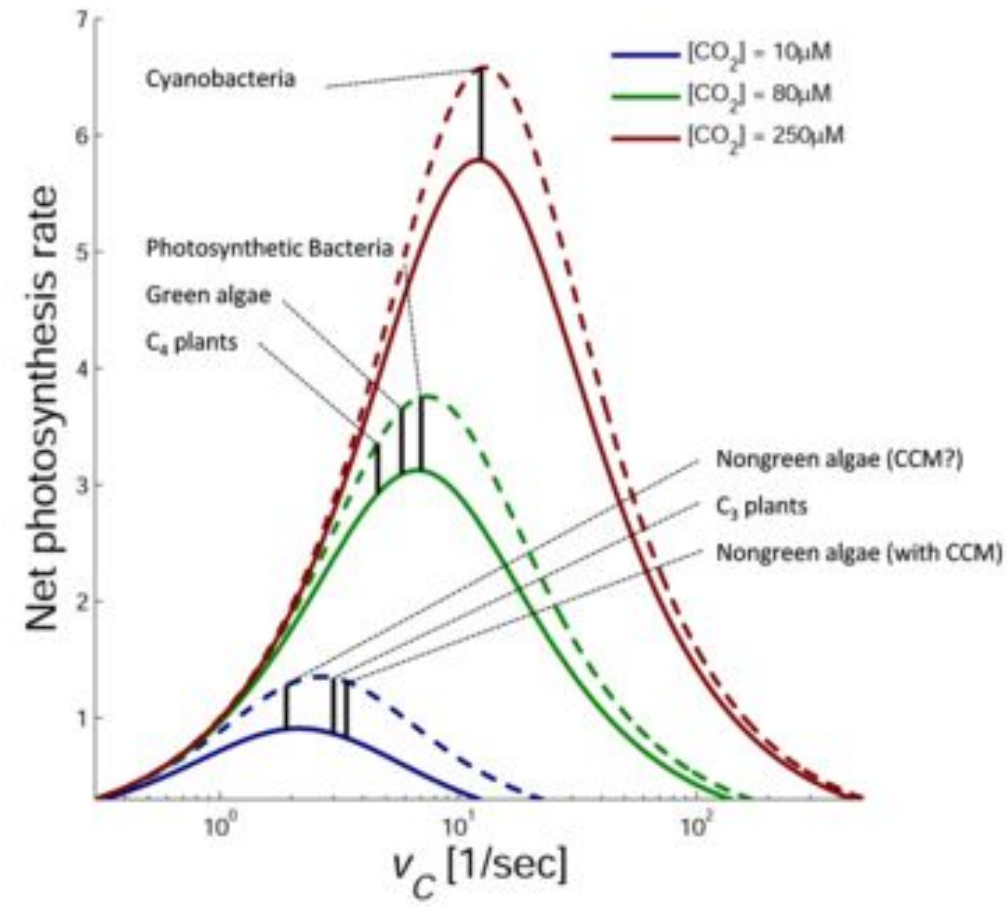

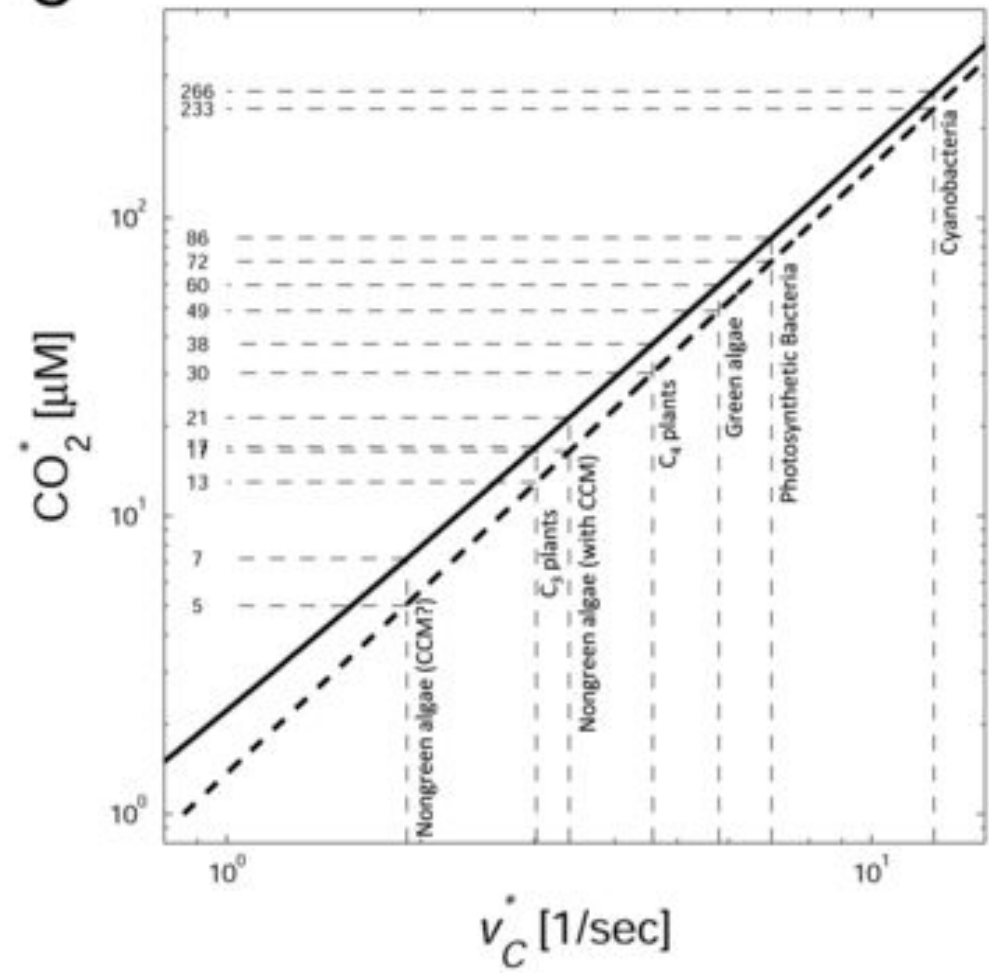

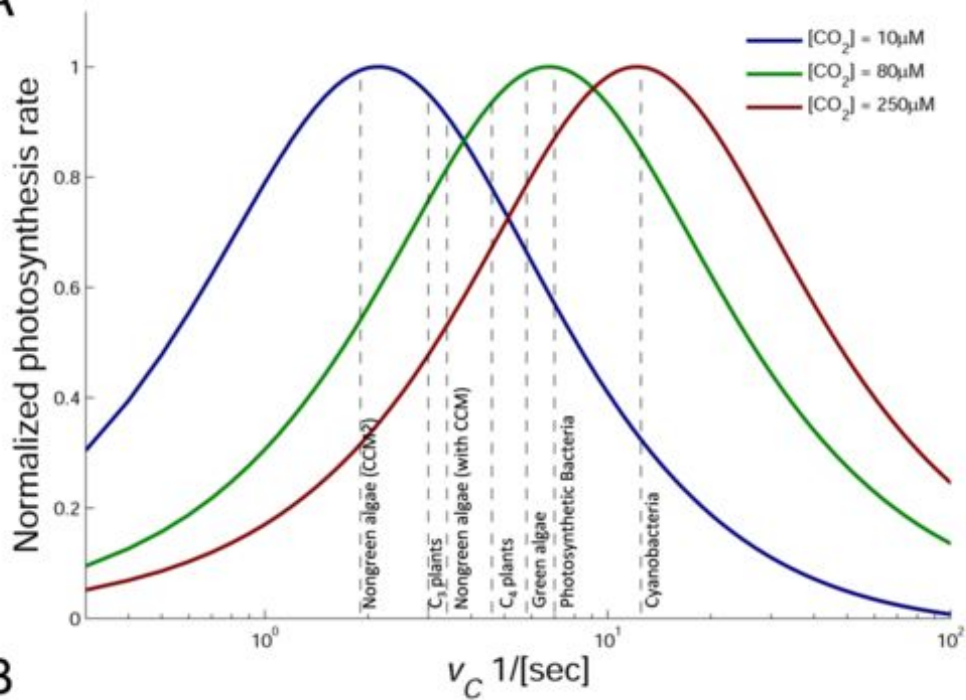

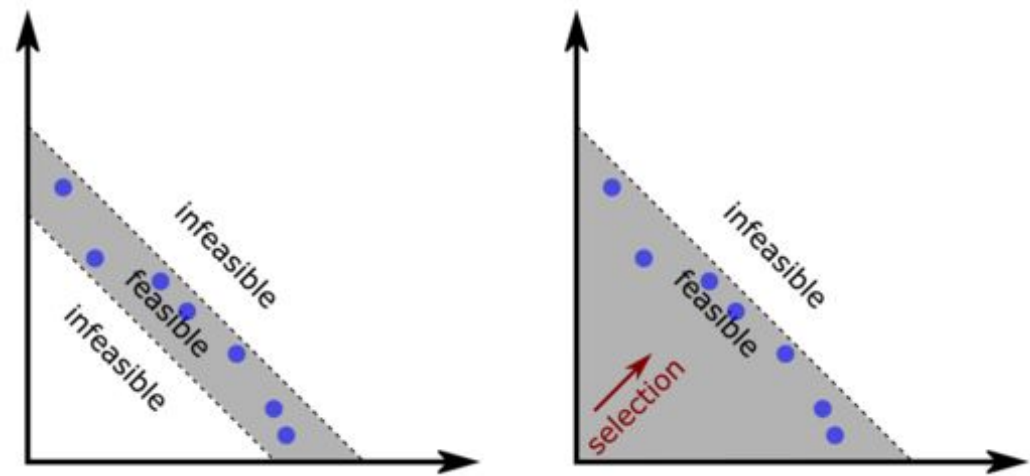